# Simultaneous Sorting of Wavelengths and Spatial Modes using Multi-Plane Light Conversion

Yuanhang Zhang[1*], He Wen[1], Alireza Fardoost[1], Shengli Fan[1], Nicolas K. Fontaine[2], Haoshuo Chen[2], Patrick L. Likamwa[1], Guifang Li[1]

[1]*CREOL, the College of Optics and Photonics, University of Central Florida, Orlando, FL 32816, USA*
[2]*Nokia Bell Labs, Crawford Hill Lab, 791 Holmdel Rd., Holmdel, NJ 07733, USA*
*Corresponding author: yuanhangzhang@knights.ucf.edu*

**We propose a wavelength-mode sorter realized by multi-plane light conversion (MPLC). For the first time, to our best knowledge, wavelengths and spatial modes can be sorted simultaneously. We first demonstrate pure wavelength sorting by a series of phase masks, which could find applications in high-power wavelength beam combining (WBC) or coarse wavelength-division multiplexing (CWDM), for example. We then present a design of a 4-wavelength, 3-mode sorter using only 5 phase masks. Insertion loss (IL) and mode dependent loss (MDL) as low as 1.27 dB and 2.45 dB can be achieved, respectively.**

## 1. Introduction

It is well known that polarization, wavelength, and spatial mode are different dimensions of light that can be used to carry information. Sorting polarization or wavelength for polarization-division multiplexing (PDM) or wavelength-division multiplexing (WDM) has long been known in the optical communication community. Recently, a high-performance mode sorter was experimentally demonstrated to sort up to 325 Laguerre-Gaussian modes [1]. A meta-optical volumetric device was recently shown to sort images into polarizations and wavelengths simultaneously [2]. However, to our best knowledge, no device has been demonstrated to sort wavelengths and spatial modes at the same time up to now. Here we propose a device that fulfills this function based on the multi-plane light conversion (MPLC) technique. MPLC consists of multiple phase masks separated by free space propagation and has been theoretically proven to have the ability to realize arbitrary unitary transformations [3]. The concept of MPLC was soon applied to many devices, including a mode sorter [1], a mode coupler [4], a mode router [5], an optical hybrid [6, 7], and for applications such as image classification [8] and object detection [9].

A schematic of the proposed device and its functionality are shown in Fig. 1. The wavelength-mode multiplexed signal from an input few-mode fiber (FMF), propagates forward through a series of phase masks. At the output, the wavelength-mode multiplexed signal is sorted into a two-dimensional (2D) array of the fundamental Gaussian beams, arranged with equal pitch. Input modes are separated into different columns (each column corresponds to one specific mode of the input), while wavelengths are separated into different rows (each row corresponds to one specific wavelength). It should be noted that wavelengths and modes are sorted in an orthogonal direction and hence each can have independent channel numbers.

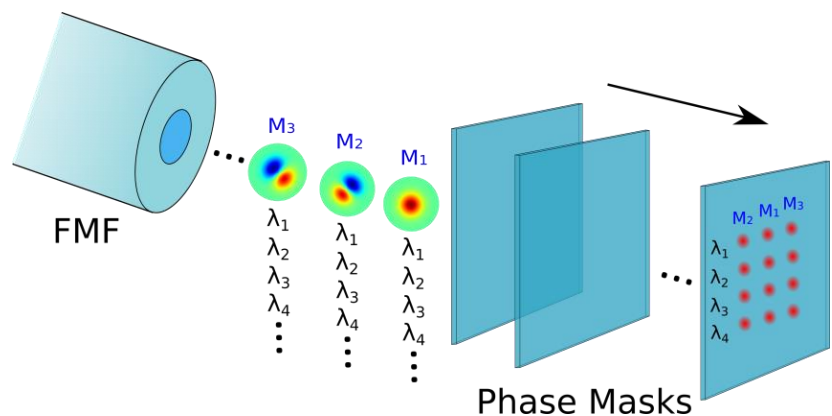

Fig. 1. Schematic of the wavelength-mode sorter. Here a 4-wavelength, 3-mode sorter demultiplexes wavelengths into 4 rows in the vertical direction, and demultiplexes modes into 3 columns in the horizontal direction. The fundamental mode $M_1$ is demultiplexed into the center column from symmetric considerations. FMF: few-mode fiber.

In section 2 we demonstrate sorting wavelengths alone by exploring the dispersive nature of a series of phase masks. In section 3, we add spatial modes and demonstrate a wavelength-mode sorter that can successfully sort 4 wavelengths and 3 spatial modes.

## 2. MPLC wavelength sorters

In demultiplexing the wavelengths, the MPLC utilizes a series of phase masks to achieve lower loss than using a simple phase grating. Fig. 2 shows the simulation results of splitting 4 wavelengths using 5 phase masks, each with 1400 ×1200 pixels. The pixel pitch is 5 μm, and the plane spacing is 18 mm. The diameter of the waist of the input Gaussian mode is 300 μm, and the diameter of the waist of the output spots is 70 μm, with a pitch of 250 μm. The input and output plane are right before the first phase mask and right after the last phase mask, respectively.

A modified wavefront-matching (WFM) algorithm was used to calculate the pattern of the phase masks [1]. This algorithm runs on a CPU (Intel Core i7-7820X) and converges in about 100 iterations, as shown in Fig. 2(c). Between phase plates we use a band-limited angular spectrum method (ASM) for free space propagation, which can effectively avoid aliasing for a long-distance propagation [10]. Once all masks are finalized, we propagate the ideal inputs at four design wavelengths but overlapped in space through the MPLC to obtain four spatially separated spots, for the 1.53 μm,1.54 μm, 1.55 μm, and 1.56 μm inputs, from top to bottom, respectively. Fig. 2(b) shows intensity profiles of output beams after the last phase mask. Fig. 2(d) plots the transmission spectrum on the output plane by calculating the ratio of actual and desired output power in each wavelength channel. Fig. 2 (e) shows that the phase masks have grating features along vertical direction, which are responsible for wavelength splitting.

We choose a pixel size of 5 μm considering the cost and fabrication capability of binary lithography, a common method for phase mask fabrication. As it's well known that the angular dispersion of a grating is inversely proportional to the pixel size, the pixel size we choose can separate wavelength with spacings as small as 10 nm. As a result, this wavelength sorter could find applications in coarse wavelength-division multiplexing (CWDM), which has a channel spacing of 20 nm.  For an actual MPLC device, it is typically implemented using reflective masks, whereby the beam bounces back and forth between a single substrate containing all the phase masks, and a flat mirror. The mask plane and mirror are usually coated with gold to reduce loss.  If this wavelength sorter is used in the reverse order, due to reciprocity, it can also be used for high-power wavelength beam combining (WBC) [11], taking advantage of the large damage threshold of gold coating.

## 3. MPLC wavelength-mode sorters

In this section we add mode demultiplexing capability to the previous mentioned wavelength sorter. As a proof-of-concept demonstration, we design a device that can sort 3 modes and 4 wavelengths simultaneously, as shown in Fig. 1.  We use the same simulation parameters as section 2, and we choose Hermite-Gaussian (HG) modes for the input modes because they are the eigenmodes of a few-mode fiber with a parabolic-index profile. The output spots have a square pitch of 250 μm, making it easy to couple into a standard 2D matrix fiber array. The fundamental mode ($HG_{00}$) is designed to map to the middle column at the output plane from symmetry considerations, to reduce the number of phase masks. It should be noted that the spacing of output spots, orientation (rows for wavelength and columns for modes), and channel number in each dimension can be adjusted according to specific needs.

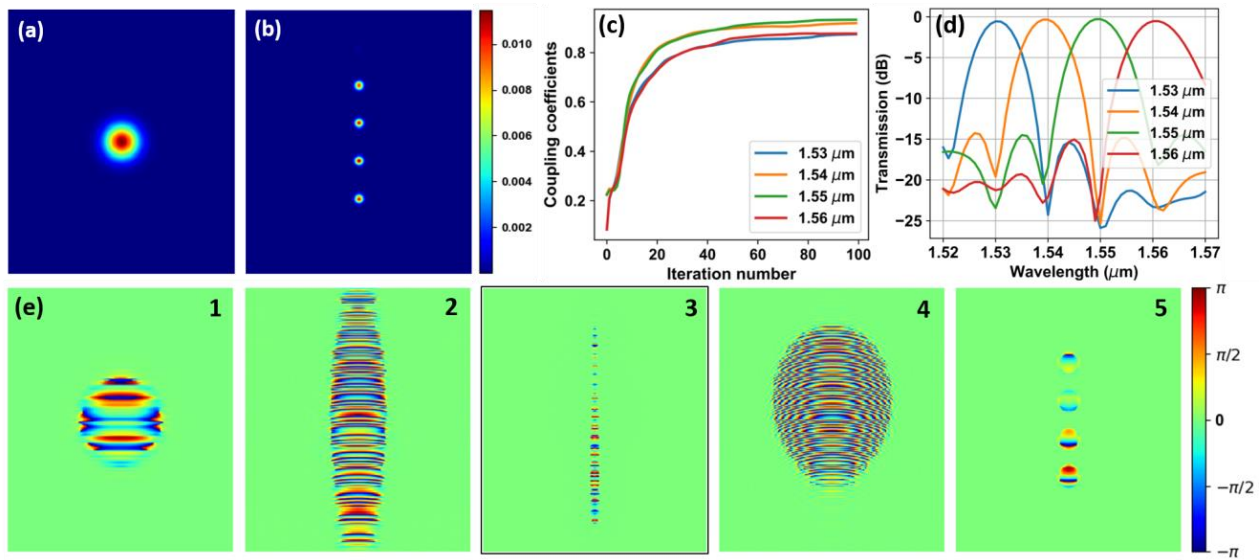

Fig. 2 Simulation results of the wavelength sorter. (a) Intensity of input $HG_{00}$ beam on the first mask. (b) intensity of actual output beams after the last mask. (c) Coupling coefficients as a function of the iteration number. (d) Transmission spectrum on the output plane. (e) 5 phase masks used to realize wavelength sorting. The grating features in the vertical direction is critical for wavelength splitting.  Note that beams in (a) and (b) are power normalized. (a), (b), and (e) all have a zoom-in ratio of 4 to clearly observe the patterns, except mask 3 (e-3), which is shown at the original size as its pattern extends almost to the edge of the entire computational window.

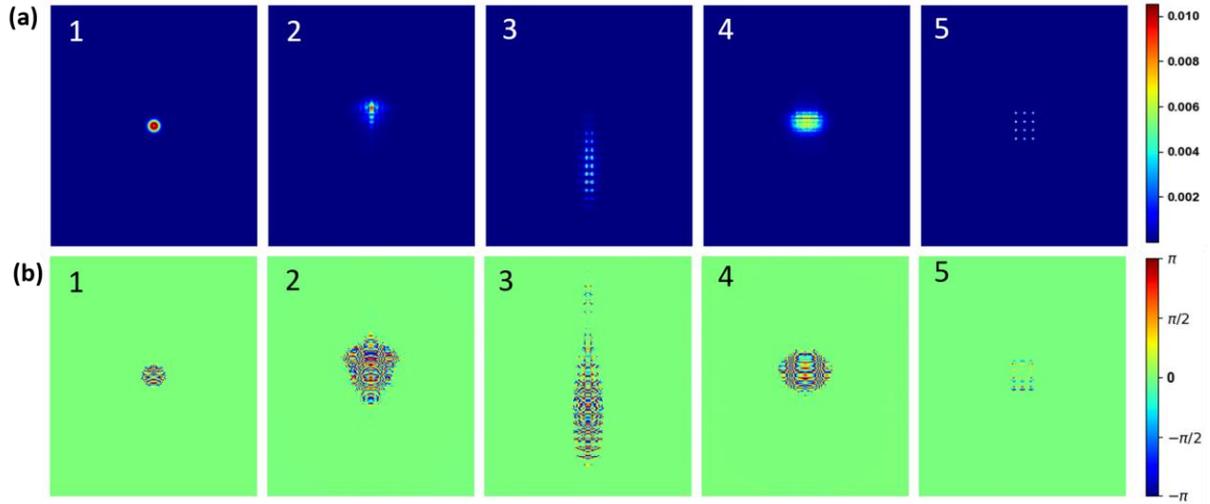

Fig. 3 Simulation results of the wavelength-mode sorter in original size of computational window. (a) The total beam intensity (power normalized) right after each mask, and (b) 5 phase masks used for wavelength-mode sorting.

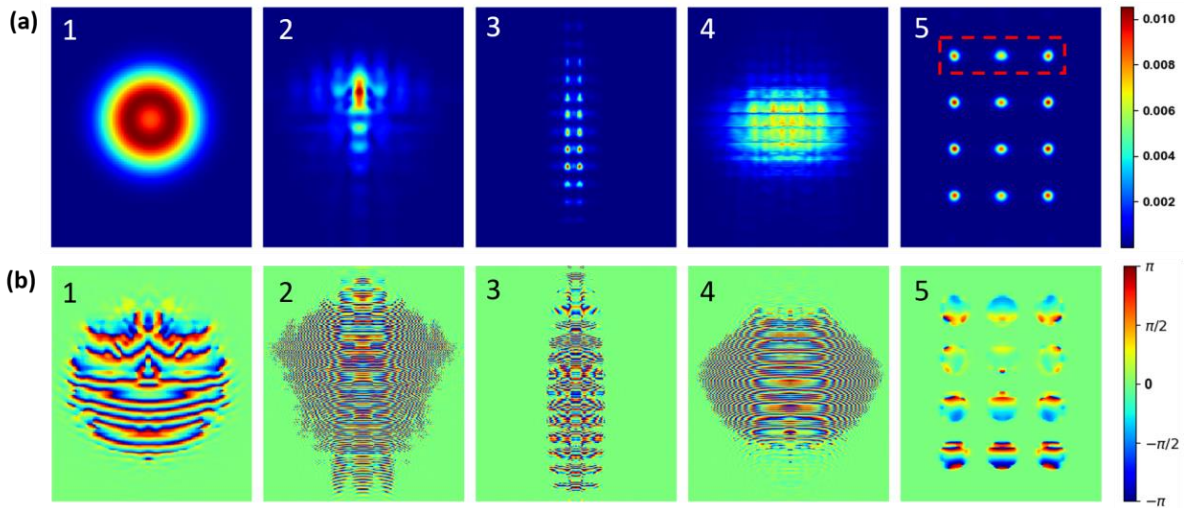

Fig. 4. Zoom-in of the region of interest (ROI) of Fig. 3 in an arbitrary ratio (each column has the same zoom-in ratio). (a) Total beam intensity (power normalized) right after each mask, and (b) 5 phase masks to realize this conversion. The red dash-line box in (a-5) indicates the region of first wavelength channel, which will be used to calculate the blue transmission curve in Fig. 7.

Fig. 3 shows simulation results of the 4-wavelength, 3-mode sorter in the original size of the computational window, while Fig. 4 are a zoom-in version of the region of interest (ROI) of Fig. 3 in an arbitrary ratio, in order to observe the beam intensity patterns and phase masks clearly.

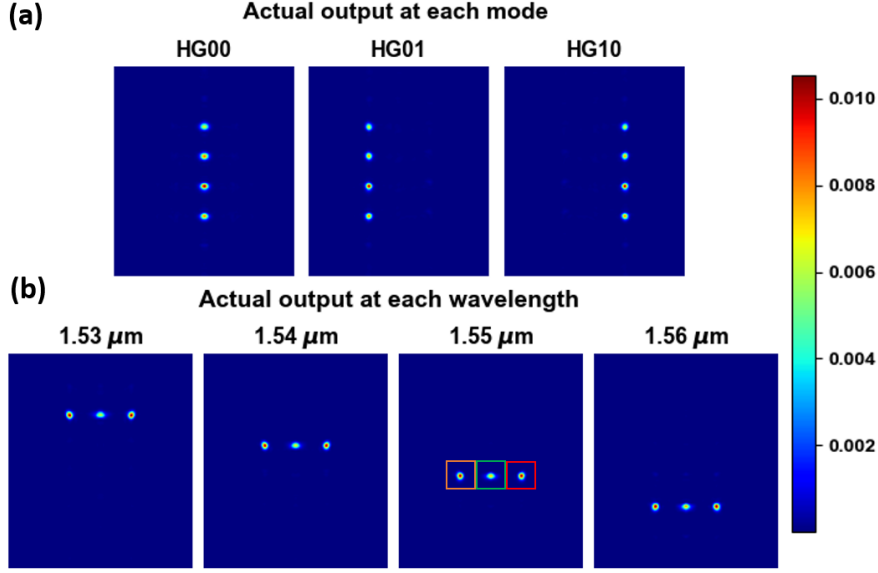

Fig. 5. Actual output intensity of this 4-wavelength, 3-mode sorter at (a) different modes and (b) different wavelengths. Beam intensity is power normalized. Both (a) and (b) has a zoom-in ratio of 4. Color-boxes in (b) at 1.55 μm are used to study the mode channel bandwidth in Fig. 8.

For this 4-wavelength, 3-mode sorter, the total independent input and output channel numbers are both 12 ($N_{mode} \times N_{wavelength}$), and every input and output are power normalized before calculating the phase distributions on the masks. The same WFM algorithm was used to optimize the phase masks. Once the masks finalized, we propagate the 12 ideal inputs through the MPLC to obtain the outputs $E_a(x, y)$, as shown in Fig. 5, which are then overlapped with the desired outputs $E_b(x, y)$ to calculate the complex coupling matrix $C$. The elements $(i, j)$ of $C$ are given by

$$C_{ij} = \frac{\iint E_{a,i}(x,y) E_{b,j}^*(x,y) dxdy}{\iint |E_{b,j}(x,y)|^2 dxdy}, \quad (1)$$

where $*$ denotes conjugation. The absolute squared values $|C_{ij}|^2$ are the power coupling coefficients, which are shown in Fig. 6(a). The overall performance of the sorter is given by extracting the insertion loss (IL) and mode-dependent loss (MDL) from the complex coupling matrix $C$ using singular value decomposition (SVD):

$$IL = \frac{1}{N} \sum_{n=1}^{N} |\eta_n|^2, \quad (2)$$

$$MDL = \frac{\max(|\eta_n|^2)}{\min(|\eta_n|^2)}, \quad (3)$$

where $\eta_n$ are the singular values of the coupling matrix $C$. For this device, IL = 1.27 dB and MDL = 2.45 dB. As the device is composed of theoretically lossless components, the insertion loss is only incurred when light is scattered into higher-order modes not supported by the system [1].

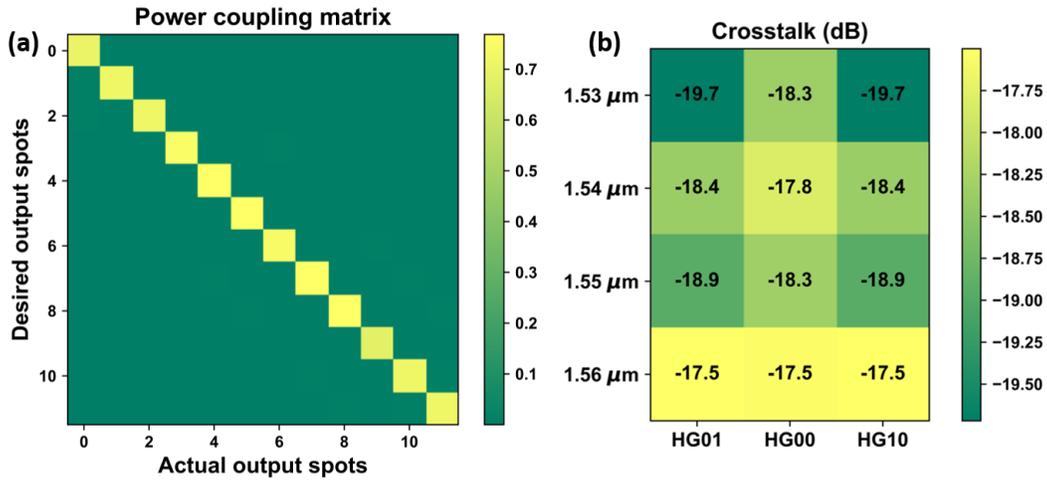

Fig. 6 Coupling matrix and crosstalk. (a) Power of the complex coupling matrix. The diagonal feature indicates this device has low crosstalk. (b) The crosstalk of each spot corresponding to every mode and wavelength at the output plane.

Fig. 6 (b) quantifies the crosstalk of each spot, which is defined as the ratio between the summation of off-diagonal elements and the diagonal element in each column of Fig. 6(a). The middle column of Fig. 6(b), corresponding to $HG_{00}$ mode, generally has a higher crosstalk than the side spots because it has two neighboring columns. The average crosstalk of 12 output spots is -18.42 dB.

We then scan the wavelength from 1.52 to 1.57 μm to study the bandwidth of each wavelength channel (channel 1 is marked by a red dash-line box in Fig.4 (a-5), for example) with results shown in Fig. 7. The transmission of each wavelength channel is the ratio between the actual output power in that boxed region to the ideal output power at the fixed design wavelength in that region. The peak value gives the insertion loss of each wavelength channel, which are 1.06 dB, 0.84 dB, 0.80 dB, 1.02 dB, respectively. Each channel has almost the same FWHM bandwidth of 9 nm.

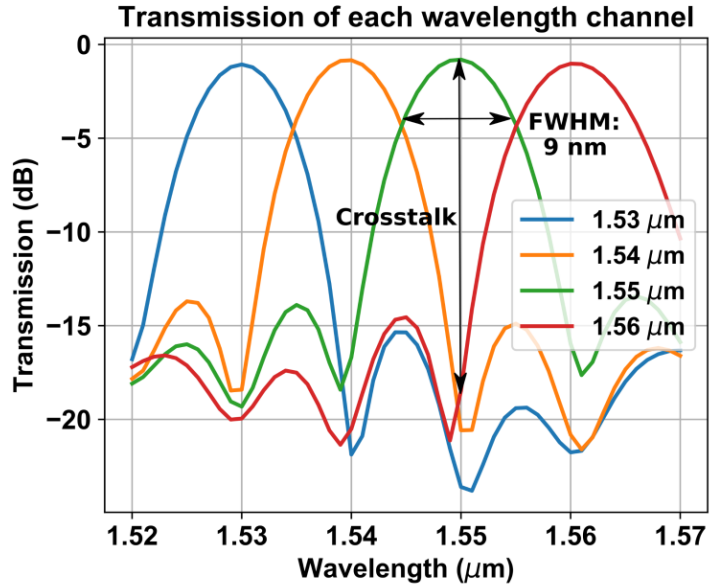

Fig. 7. Transmission spectrum of each wavelength channel. The transmission of each channel is the ratio between the actual output power in that boxed region to the ideal output power at the fixed design wavelength in each box region. For example, channel 1 is shown by a red dash-line box in Fig. 4(a-5). FWHM: full width at half maximum.

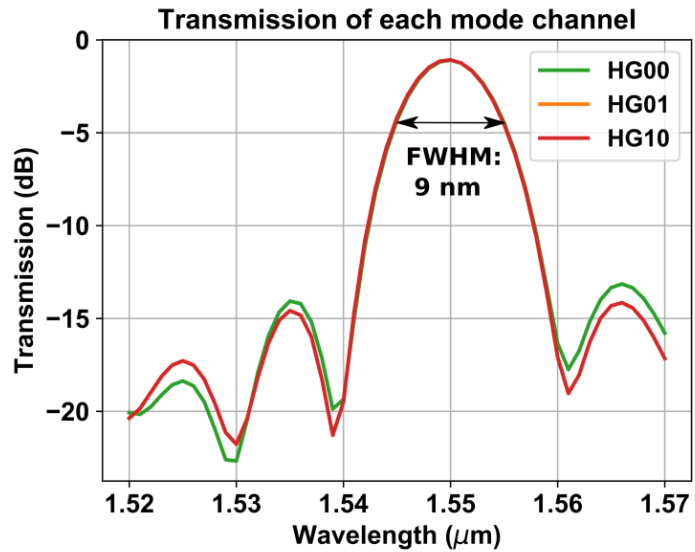

Fig. 8 Transmission spectrum of each mode channel. The transmission of each mode channel, corresponding to the 1.55 μm channel, is calculated by the ratio of actual output power and desired output power in each square region shown by the color-boxes in Fig. 5(b).

In Fig. 8 we present the bandwidth of output mode channel by plotting the transmission as a function of wavelength. Modes from the third row (1.55 μm channel) on the output plane are used in the calculation, as shown by the color-boxes in Fig. 5(b). The transmission of each mode channel is the ratio between the actual output power in that boxed region to the ideal output power at the fixed design wavelength in that region. Three curves are almost overlap, indicating they are in the same wavelength channel. Each mode channel also has a FWHM bandwidth of 9 nm.

## 4. Conclusion

In conclusion, we design a high-performance wavelength-mode sorter using the multi-plane light conversion technique. This device could potentially be used in a coarse wavelength division multiplexed (CWDM) system together with spatial division multiplexing (SDM). We also design a wavelength sorter using a series of phase masks, with could find applications in high-power wavelength beam combining (WBC).

**Funding.** Army Research Office (W911NF1710553, W911NF1810365, W911NF1910385); National Science Foundation (ECCS-1808976, ECCS-1932858); Office of Naval Research (N00014-20-1-2441).

**Acknowledgment**. Y. Zhang acknowledges the financial support from the China Scholarship Council (CSC) Scholarship No. 201606250006.